\documentclass[11pt]{article}

\usepackage{mathtools}
\usepackage{amssymb}
\usepackage{dsfont}
\usepackage{slashed}
\usepackage{fullpage}
\usepackage[colorlinks=true,linkcolor=blue,citecolor=magenta,linktocpage=true]{hyperref}
\usepackage{titlesec}
\titleformat*{\section}{\normalsize\bfseries}
\titleformat*{\subsection}{\normalsize\bfseries}
\titleformat*{\subsubsection}{\normalsize\bfseries}
\usepackage{cite}

\DeclareMathAlphabet{\bbvar}{U}{BOONDOX-ds}{m}{n}

\makeatletter
\renewcommand{\@dotsep}{10000}
\makeatother

\def\be#1\ee{\begin{align}#1\end{align}}
\def\nn{\nonumber}
\def\q{\qquad}
\def\f{\frac}
\def\eps{\varepsilon}
\def\teps{\tilde{\varepsilon}}
\def\lb{\big\lbrace}
\def\rb{\big\rbrace}

\def\ip{\lrcorner\,}
\def\ipp{\ip\!\!\!\ip}

\def\as{{\ast}}
\def\per{\text{\tiny{$\perp$}}}
\def\para{\text{\tiny{$\parallel$}}}

\def\bb#1{\mathbb{#1}}

\def\Tr{\mathrm{Tr}}

\def\De{\mathrm{D}}
\def\de{\mathrm{d}}

\def\C{\mathcal{C}}
\def\D{\mathcal{D}}

\def\F{\mathcal{F}}
\def\G{\mathcal{G}}

\def\L{\mathcal{L}}
\def\N{\mathcal{N}}

\def\Q{\mathcal{Q}}

\def\V{\mathcal{V}}
\def\X{\mathcal{X}}

\numberwithin{equation}{section}

\begin{document}

\title{\Large{\textbf{\sffamily Lorentz-diffeomorphism edge modes in 3d gravity}}}
\author{\sffamily Marc Geiller}
\date{\small{\textit{Perimeter Institute for Theoretical Physics,\\ 31 Caroline Street North, Waterloo, Ontario, Canada N2L 2Y5}}}

\maketitle

\begin{abstract}
The proper definition of subsystems in gauge theory and gravity requires an extension of the local phase space by including edge mode fields. Their role is on the one hand to restore gauge invariance with respect to gauge transformations supported on the boundary, and on the other hand to parametrize the largest set of boundary symmetries which can arise if both the gauge parameters and the dynamical fields are unconstrained at the boundary. In this work we construct the extended phase space for three-dimensional gravity in first order connection and triad variables. There, the edge mode fields consist of a choice of coordinate frame on the boundary and a choice of Lorentz frame on the bundle, which together constitute the Lorentz-diffeomorphism edge modes. After constructing the extended symplectic structure and proving its gauge invariance, we study the boundary symmetries and the integrability of their generators. We find that the infinite-dimensional algebra of boundary symmetries with first order variables is the same as that with metric variables, and explain how this can be traced back to the expressions for the diffeomorphism Noether charge in both formulations. This concludes the study of extended phase spaces and edge modes in three-dimensional gravity, which was done previously by the author in the BF and Chern--Simons formulations.
\end{abstract}

\thispagestyle{empty}
\newpage
\setcounter{page}{1}

\hrule
\tableofcontents
\vspace{0.7cm}
\hrule


\section{Motivations}

In gauge theory and gravity, several technical and conceptual issues arise when trying to make sense out of local subregions of spatial slices. In the quantum theory for example, the Hilbert space of physical states for the union of two neighboring subregions does not factorize spatially due to the presence of constraints and to the non-locality of gauge-invariant observables \cite{Buividovich:2008gq,Donnelly:2011hn}. This prevents entanglement entropy between the subregions from being defined as the von Neumann entropy of a reduced density matrix, since the very definition of this latter relies on the existence of a factorized Hilbert space. At the classical level on the other hand, gauge transformations fail to be null directions of the symplectic structure unless some conditions are imposed on the gauge parameters or the dynamical fields at the boundary of the local subregion, and, similarly, boundary conditions are also required in order to construct quasi-local gauge-invariant observables. However, if the local subregion and its boundary (or entangling surface) are arbitrary, one cannot impose such conditions since they would amount to freezing the dynamics of the theory.

Interestingly, these quantum and classical puzzles can be resolved in one go by introducing boundary degrees of freedom in the form of edge mode fields \cite{Donnelly:2016auv}. This leads in the quantum theory to the picture of an extended Hilbert space, which has successfully been used in computations of entanglement entropy in (continuum and lattice) gauge theory \cite{Buividovich:2008gq,Donnelly:2011hn,Casini:2013rba,Casini:2014aia,Radicevic:2014kqa,Donnelly:2014fua,Donnelly:2014gva,Hung:2015fla,Ghosh:2015iwa,Aoki:2015bsa,Donnelly:2015hxa,Delcamp:2016eya,Fliss:2017wop}. In a classical context, the analogue of this construction is that of an extended phase space. On this extended phase space, the symplectic structure has a boundary contribution containing new physical edge mode fields whose role is to restore the gauge invariance which is broken by the presence of the boundary, very much in the spirit of Stueckelberg fields \cite{Ruegg:2003ps}. There is a minimalistic and systematic way of introducing this extended symplectic structure in any gauge field theory \cite{Donnelly:2016auv,Geiller:2017xad,Speranza:2017gxd}. The advantage of this construction is that it puts forward a clear distinction between gauge transformations and symmetries. Gauge transformations are non-physical and have a vanishing on-shell Hamiltonian charge even when the gauge parameters have non-trivial support on the boundary. As mentioned earlier, it is important to allow for arbitrary gauge parameters and field configurations on the boundary because this latter is at an arbitrary location and may simply represent a fictitious entangling surface. Symmetries, on the other hand, are physical transformations which map between different physical states. Generators of symmetries are given by a non-vanishing Hamiltonian boundary charge and satisfy a boundary symmetry algebra. This latter encodes in a precise sense the symmetries which govern the gluing of subregions, either as classical phase spaces or as quantum Hilbert spaces \cite{Donnelly:2016auv}.

The existence of boundary symmetries and degrees of freedom in gauge field theory is of course not a new topic \cite{MR974271,MR1025431,Brown:1986nw,Carlip:1994gy,Balachandran:1995qa,Coussaert:1995zp,Grumiller:2016pqb,Grumiller:2017sjh}. One natural question is therefore to understand how the boundary symmetry algebras and their generators on the extended phase space are related to the boundary symmetries and observables which have previously been constructed in e.g. general relativity \cite{Regge:1974zd,Balachandran:1994up,Balachandran:1995qa,Husain:1997fm,Szabados:2005wi,Wieland:2017ksn}. This was answered in \cite{Geiller:2017xad}, where it was shown that the generators of the boundary symmetries on the extended phase space are simply a ``dressed'' version (in a sense which will be explicit below) of the usual Hamiltonian boundary charges\footnote{The Hamiltonian boundary charges are not simply given by the Noether charges, but may contain an additional piece (see e.g. equation (2.32) in \cite{Geiller:2017xad}).} (which we review in appendix \ref{app:usual}) associated with gauge transformations. While the usual Hamiltonian boundary charges can be promoted to observables if the constraints are defined with smearing fields (i.e. gauge parameters)  which are vanishing on the boundary, the extended phase space construction provides a way of relaxing this requirement, and of constructing boundary observables without imposing any boundary conditions on the gauge parameters or the dynamical fields. In other words, the boundary symmetries and observables on the extended phase space correspond to the ``maximal'' amount of physical symmetries which can be obtained if all the fields and gauge parameters are unconstrained on the boundary.

The extended phase space and boundary symmetries were constructed in \cite{Donnelly:2016auv} in the case of non-Abelian Yang--Mills theory and gravity in metric variables. It was shown that the boundary symmetries of gravity in a local subregion are given by the algebra of $\text{Diff}(S)\ltimes\text{SL}(2,\mathbb{R})$, together with possible central extensions studied in \cite{Speranza:2017gxd}, where $S$ is the co-dimension 2 boundary (and is the only dimension-dependent object). The goal of the present work is to show that the same algebra is obtained in three-dimensional gravity with first order connection and triad variables. This can come as a surprise, since first order gravity possesses internal Lorentz freedom in addition to diffeomorphism, and one could therefore have expected that ``more'' gauge transformations can become physical on the boundary. As we will see, this is not the case because the Noether charge for diffeomorphisms in first order variables differs importantly from the Noether charge in metric variables in that it does not depend on derivatives of (and is linear in) the vector field. For this reason, diffeomorphisms which have a vanishing generating vector field (but with non-vanishing derivative) have a vanishing Noether charge in first order variables but not in metric variables. However, such diffeomorphisms which do not move points do actually define local Lorentz transformations (essentially by virtue of the equivalence principle), which is precisely the information encoded in the charges for Lorentz gauge transformations in the first order formulation, thereby explaining why there is the same physical information encoded in the Noether charges of first order and metric gravity\footnote{This fact can also be understood in light of the calculation presented in appendix \ref{appendix:GHY}.}. Consequences of this fact were also studied before in the context of black hole entropy as Noether charge \cite{Jacobson:2015uqa}. Here we wish to explain how the first order diffeomorphism and Lorentz Noether charges are used to construct the extended symplectic structure and to define the boundary observables and their algebra. This will serve as a simple and hopefully pedagogical example illustrating this generic construction.

In the first part of this work, we will introduce the extended phase space structure. For this, we will explain how to include edge mode fields associated with diffeomorphism and Lorentz transformations based on the requirement of gauge invariance of the symplectic potential. Then, we will simply study how gauge transformations act on this extended phase space, and show that their generators are vanishing on-shell and therefore do not posses a Hamiltonian charge. Finally, we will turn to boundary symmetries. After defining how these symmetries act on the variables of the extended phase space, we will compute their Hamiltonian generators, study their integrability, and show that the algebra of boundary symmetries is that of $\text{Diff}(S)\ltimes\text{SL}(2,\mathbb{R})$ together with possible central extensions, in agreement with the results of \cite{Donnelly:2016auv,Speranza:2017gxd} derived in the metric formulation.

\section{First order gravity}

We are interested in first order three-dimensional gravity with a cosmological constant of arbitrary sign. The Lagrangian for this theory is given by
\be\label{Lagrangian}
L=e\wedge\left(F+\f{1}{6\ell^2}[e\wedge e]\right),
\ee
where $F$ is the field strength of the connection $\omega$, and a trace in the appropriate Lie algebra is understood\footnote{In the Lorentzian case, this Lie algebra is that of the three-dimensional Lorentz group SO(2,1) (or its double cover SU(1,1)), while in the Euclidean case it is the algebra of SO(3) (or its double cover SU(2)). The fields entering the Lagrangian are Lie algebra-valued one-forms, e.g. $e=e^i_\mu J_i\de x^\mu$, with $J_i$ a basis of the algebra, and the trace can simply be chosen up to normalization to be $\Tr(J_i,J_j)=\delta_{ij}$.}. The equations of motion are the torsion-free condition and the curvature constraint, i.e.
\be
\De e=0,\q F_\ell\coloneqq F+\f{1}{2\ell^2}[e\wedge e]=0,
\ee
from which one can see that the Lagrangian is not vanishing on-shell (unless $\ell^2=\infty$). This property is of course true in any dimension for gravity with a cosmological constant (whether formulated in first order or metric variables), and also for non-Abelian Chern--Simons theory. We will see later on that this fact will result in an additional contribution in the extended symplectic structure, which however will only play a role when discussing the central extension arising from diffeomorphisms that move the boundary $S$.

Three-dimensional gravity, as defined by \eqref{Lagrangian}, is invariant under three types of transformations. At the infinitesimal level, they are given by the internal Lorentz transformations
\be\label{infinitesimal Lorentz}
\delta_\alpha e=[e,\alpha],\q\delta_\alpha\omega=\De\alpha,
\ee
the so-called topological translations
\be\label{infinitesimal translations}
\delta_\phi e=\De\phi,\q\delta_\phi\omega=\f{1}{\ell^2}[e,\phi],
\ee
and the diffeomorphisms
\be\label{infinitesimal diffeo}
\L_\xi e=\de(\xi\ip e)+\xi\ip(\de e),\q\L_\xi\omega=\de(\xi\ip\omega)+\xi\ip(\de\omega).
\ee
These transformations are not independent, as diffeomorphisms can be realized on-shell as a combination of field dependent Lorentz transformations and translations \cite{MR974271,Geiller:2017xad}. In order to describe the edge modes and the extended phase space, one therefore has to first choose a set of independent gauge transformations. In \cite{Geiller:2017xad} we have chosen the Lorentz transformations and the translations. The goal of the present paper is to extend this construction to the case of Lorentz transformations and diffeomorphisms.

\subsection{Extended phase space}

Let us start by focusing on diffeomorphisms. As explained in \cite{Donnelly:2016auv,Speranza:2017gxd}, in order to describe the edge modes associated with diffeomorphism transformations, the field which one has to introduce is a map $X$ from an open subset of $\bb{R}^3$ to the three-dimensional spacetime manifold $M$. This can essentially be thought of as a choice of coordinates. In particular, it will be important to think of local spatial subregions $\Sigma$ as being defined as the image under $X$ of some $\sigma\subset\bb{R}^3$. We will therefore denote the region by $\Sigma=X(\sigma)$, its boundary by $S=X(s)$, and assume for the sake of simplicity that $\Sigma$ can be covered by a single choice of $X$.

Following \cite{Speranza:2017gxd}, the most straightforward way of introducing this new field in a given theory is to simply substitute its Lagrangian $L$ for the pullback $X^*L$. If we denote the collection of initial fields of the theory by $\Phi$, by virtue of diffeomorphism invariance we then have that $X^*L[\Phi]=L[X^*\Phi]$. Let us now see which pre-symplectic potential is obtained by computing the functional variation of this Lagrangian while treating $X$ as an additional dynamical variable. Denoting the equations of motion by $E$, we have
\be
\delta(L[X^*\Phi])=E[X^*\Phi]\wedge\delta(X^*\Phi)+\de\theta[X^*\Phi,\delta(X^*\Phi)],
\ee
and the total exterior derivative term, identified as the potential, can be rewritten as\footnote{We denote by $\ip$ the contraction in spacetime of a vector field and a differential form, which in components corresponds to e.g. $\xi\ip e=\xi^\mu e_\mu$ or $\big(\xi\ip(\de e)\big)_\nu=\xi^\mu(\partial_\mu e_\nu-\partial_\nu e_\mu)$. By extension, we also denote by $\ipp$ the contraction in field space of a tangent vector and a field space differential form. With this useful notation, gauge transformations $\delta_\alpha$ or diffeomorphisms $\L_\xi$ can be seen as tangent vectors on field space, and therefore contracted with variational expressions such as e.g. the field space two-form $\delta\omega\wedge\delta e=\delta_1\omega\wedge\delta_2e-\delta_2\omega\wedge\delta_1e$, to obtain a field space one form $\delta_\alpha\ipp(\delta\omega\wedge\delta e)=\delta_\alpha\omega\wedge\delta_1e-\delta_1\omega\wedge\delta_\alpha e$, where $\delta_\alpha\omega$ and $\delta_\alpha e$ are then understood as infinitesimal gauge transformations. In \cite{Donnelly:2016auv,Speranza:2017gxd}, the contractions $\ip$ and $\ipp$ are denoted respectively by $i$ and $I$.}
\be\label{framed potential}
\theta[X^*\Phi,\delta(X^*\Phi)]=X^*(\theta[\Phi,\delta\Phi]+\theta[\Phi,\L_\X\Phi])=X^*(\theta+\L_\X\ipp\theta).
\ee
This equality follows from the commutation property \cite{Donnelly:2016auv,Speranza:2017gxd}
\be\label{commutation property}
\delta(X^*\Phi)=X^*(\delta\Phi+\L_\X\Phi),
\ee
where the object $\X\coloneqq\delta X\circ X^{-1}$, which is a vector field on $M$ and also a 1-form in field space, appears only since we consider that $\delta X\neq0$, or in other words that $X$ is part of the space of fields of the theory. The potential \eqref{framed potential} has the desirable property of being invariant under arbitrary finite and possibly field-dependent diffeomorphisms (and therefore also under their infinitesimal version), which is not the case of the ``bare'' potential $\theta=\theta[\Phi,\delta\Phi]$ \cite{Speranza:2017gxd}. Let us now proceed with an important rewriting.

Recall that a spacetime diffeomorphism parametrized by a vector field $\xi$ acts on the Lagrangian as $\L_\xi L=\de(\xi\ip L)$, and can be assigned a Noether current given by\footnote{In the rest of the text $\simeq$ will refer to an equality which holds on-shell, i.e. using the equations of motion.}
\be
J=\L_\xi\ipp\theta-\xi\ip L\simeq\de Q[\xi],
\ee
where the on-shell quantity $Q[\xi]$ is the Noether charge associated with a diffeomorphism vector field $\xi$ \cite{Iyer:1994ys}. For the Lagrangian \eqref{Lagrangian}, whose potential is $\theta=\delta\omega\wedge e$, this Noether charge is explicitly given by
\be
J
&=\L_\xi\omega\wedge e-\xi\ip L\nn\\
&=\De(\xi\ip\omega)\wedge e+\xi\ip F\wedge e-\xi\ip L\nn\\
&=\De(\xi\ip\omega)\wedge e-\xi\ip eF_\ell\nn\\
&=\de(\xi\ip\omega e)-\xi\ip\omega\De e-\xi\ip eF_\ell\nn\\
&\simeq\de(\xi\ip\omega e).
\ee
In this succession of equalities, we have used the fact that the Lie derivative of a connection can be written in terms of the associated covariant derivative and curvature as
\be
\L_\xi\omega=\de(\xi\ip\omega)+\xi\ip(\de\omega)=\De(\xi\ip\omega)+\xi\ip F,
\ee
then made use of the identity
\be
\xi\ip L=\xi\ip eF_\ell+\xi\ip F\wedge e,
\ee
and finally integrated by parts before using the equations of motion to obtain a total exterior derivative. As mentioned above, the resulting quantity $Q[\xi]=\xi\ip\omega e$ is the Noether charge associated to a diffeomorphism in three-dimensional first order gravity. With this, we can now integrate \eqref{framed potential} over $\sigma$ and rewrite the resulting quantity as
\be
\int_\sigma X^*(\theta+\L_\X\ipp\theta)=\int_{X(\sigma)}\theta+\L_\X\ipp\theta\simeq\int_\Sigma\theta+\X\ip L+\int_SQ[\X],
\ee
where we should keep in mind from now on that $\Sigma=X(\sigma)$ and $s=X(s)$. Note that this new potential has a boundary contribution which is given by the Noether charge associated with the variational vector field $\X$, and also a bulk contribution from $\X$ in the case of a theory whose Lagrangian is not vanishing on-shell.

We now have to account for the Lorentz transformations as well. This can be done by introducing a group-valued function $u$ transforming under finite Lorentz transformations with parameter $h$ as $h^*u=h^{-1}u$. The potential which incorporates this new field is then given by \cite{Geiller:2017xad}
\be\label{extended potential}
\Theta
&=\int_\Sigma\theta+\X\ip L+\int_S\X\ip\omega e+(\delta uu^{-1}+\X\ip\de uu^{-1})e\nn\\
&=\int_\Sigma\theta+\X\ip L+\int_SQ[\X]+Q[\delta uu^{-1}+\X\ip\de uu^{-1}],
\ee
where in the second line we have introduced the Noether charge\footnote{Note that for simplicity of notation we denote by $Q[\,\cdot\,]$ the Noether charges for both diffeomorphisms and Lorentz transformations. What distinguishes them is their argument, which for the former is a vector field and for the latter a Lie algebra element.} $Q[\alpha]=\alpha e$ associated with Lorentz transformations with Lie algebra parameter $\alpha$. The group element $u$ is the field introduced in order to describe the edge modes associated with Lorentz transformations, and one can see that it gives two contributions. The first one can be understood as making the potential invariant under Lorentz transformations, while the second one is a coupling to the diffeomorphism edge mode field $X$ which is necessary in order to make the potential invariant under both types of transformations.

Note that while in \cite{Geiller:2017xad} the form of the potential incorporating the edge mode field $u$ was derived based on the requirement of gauge-invariance under finite transformations $h^*$, one can also obtain this result following what has been done above for diffeomorphisms: one can simply replace the Lagrangian $L[\Phi]$ by $L[u^*\Phi]$, and then inspect the resulting potential $\theta[u^*\Phi,\delta(u^*\Phi)]$ to read off the contribution of $u$. This shows that the method of \cite{Speranza:2017gxd} is actually a generic way of obtaining the edge mode contribution for any type of gauge transformations.

We can now write the statement of gauge invariance for the extended potential \eqref{extended potential}. Denoting the action of a finite gauge transformation by $\Theta^\epsilon\coloneqq\Theta[\epsilon^*\Phi,\delta(\epsilon^*\Phi)]$, we have
\be\label{finite transformations of extended potential}
\Theta^Y\simeq\Theta\simeq\Theta^h,
\ee
where $Y$ is a finite (and possibly field-dependent) diffeomorphism acting by pull-back and $h$ a group element parametrizing a finite (and possibly field-dependent) Lorentz transformation. This invariance under finite gauge transformations implies of course the invariance under their infinitesimal version. At the infinitesimal level, invariance can be shown by using the explicit action of the gauge transformations on the various fields, which we will need later on. In addition to \eqref{infinitesimal Lorentz} and \eqref{infinitesimal diffeo}, this is given by the transformation rules
\be\label{transformation of extra fields}
\delta_\alpha u=-\alpha u,\q\delta_\alpha X=0,\q\L_\xi u=\xi\ip\de u,\q\L_\xi\ipp\X=\L_\xi X\circ X^{-1}=-\xi.
\ee
With this we can then show that
\be\label{infinitesimal transformations of extended potential}
\L_\xi\ipp\Theta\simeq0\simeq\delta_\alpha\ipp\Theta.
\ee
In other words, we have constructed an extended potential $\Theta$ which contains two additional fields, $X$ and $u$, whose role is to restore on-shell gauge invariance in the sense of \eqref{finite transformations of extended potential} and \eqref{infinitesimal transformations of extended potential}.

What we are really aiming for is the pre-symplectic 2-form, which is obtained by computing the field space variation of the potential. For this, we have to remember that $\Sigma$ and $S$ are defined through $X$, which can also be acted on by the variations. When commuting the variation and the integration over a region, we therefore have to use the commutation property \eqref{commutation property} to write
\be\label{commuting variation and integral}
\delta\int_\Sigma f=\delta\int_\sigma X^*f=\int_\sigma\delta(X^*f)=\int_\sigma X^*(\delta f+\L_\X f)=\int_\Sigma\delta f+\L_\X f.
\ee
Let us consider separately the bulk and boundary contributions to the potential \eqref{extended potential}. Varying the bulk part leads to
\be
\delta\Theta_\Sigma
&=\int_\Sigma\delta\theta+\delta(\X\ip L)+\L_\X\theta+\L_\X(\X\ip L)\nn\\
&=\int_\Sigma\delta\theta+\delta\X\ip L-\X\ip\delta L+\L_\X\theta+\L_\X(\X\ip L)\nn\\
&=\int_\Sigma\delta\theta+\delta\X\ip L-\X\ip\delta L+\X\ip(\de\theta)+\L_\X(\X\ip L)+\int_S\X\ip\theta\nn\\
&\simeq\int_\Sigma\delta\theta+\delta\X\ip L+\L_\X(\X\ip L)+\int_S\X\ip\theta,
\ee
where we have first used the Leibniz rule for the variation, then the Cartan formula for the Lie derivative, and finally the fact that
\be
\delta L-\de\theta=\text{EOMs}\wedge\delta\Phi\simeq0.
\ee
Let us now show that the bulk contributions in addition to $\delta\theta$ actually combine in a total exterior derivative (note that they simply disappear in the case of theories whose Lagrangian is vanishing on-shell). The proof uses the expression for the variation of $\X$, which is \cite{Donnelly:2016auv,Speranza:2017gxd}
\be
\delta\X=-\f{1}{2}[\X,\X],
\ee
as well as the identity $[\xi,\zeta]\ip=[\L_\xi,\zeta\ip]$ for the Lie derivative and vector fields, which here has to be used carefully since the $\X$'s are anti-commuting variational vector fields. Combining these formulas, one can show that
\be
\delta\X\ip L+\L_\X(\X\ip L)=\f{1}{2}\de(\X\ip\X\ip L).
\ee
We therefore get the result that
\be
\delta\Theta_\Sigma=\int_\Sigma\delta\theta+\int_S\X\ip\theta+\f{1}{2}\X\ip\X\ip L.
\ee
Combining this with the variation $\delta\Theta_S$, we finally get that the symplectic 2-form is given by
\be\label{extended 2-form 1}
\Omega=\Omega_\Sigma+\Omega_S=\int_\Sigma\delta\theta+\int_S\X\ip\theta+\f{1}{2}\X\ip\X\ip L+(\delta+\X\ip\de)(Q[\X]+Q[\delta uu^{-1}+\X\ip\de uu^{-1}]).
\ee
We have chosen to write this compact and general expression since it enables us to recover the extended symplectic structure of metric gravity as well. For this, one must simply set $u$ to the identity, and use the expression $Q[\xi]=\as\,\de(g\xi)=\eps_{\mu\nu}\nabla^\mu\xi^\nu$ for the diffeomorphism Noether charge in metric variables \cite{Donnelly:2016auv,Speranza:2017gxd}. For the calculations which will follow, one can however be more explicit. Using
\be
\delta(\delta uu^{-1})=\f{1}{2}[\delta uu^{-1},\delta uu^{-1}],\q\delta(\de uu^{-1})=u\de(u^{-1}\delta u)u^{-1},
\ee
one can show that
\be\label{extended 2-form 2}
&\phantom{=\ }(\delta+\X\ip\de)(Q[\X]+Q[\delta uu^{-1}+\X\ip\de uu^{-1}])\nn\\
&=-\f{1}{2}\big([\X,\X]\ip(\omega+\de uu^{-1})-[\delta uu^{-1},\delta uu^{-1}]+2\X\ip u\de(u^{-1}\delta u)u^{-1}\big)e\nn\\
&\phantom{=\ }-\X\ip\delta\omega e-(\X\ip\omega+\delta uu^{-1}+\X\ip\de uu^{-1})\delta e\nn\\
&\phantom{=\ }+\X\ip\big(\de(\X\ip\omega e+\delta uu^{-1}e+\X\ip\de uu^{-1}e)\big).
\ee
The extended symplectic structure given by \eqref{extended 2-form 1} and \eqref{extended 2-form 2} is the main result of this subsection. It defines the extended phase space containing the new fields $X$ and $u$ which have the role of restoring gauge invariance. One can see as expected that the bulk symplectic structure is unchanged, and that the new fields contribute only at the boundary\footnote{Although $X$ appears implicitly in the definition of $\Sigma$ and $S$, what really matters is the actual variation of $X$, which is contained in $\X$ and indeed only supported at the boundary.}. We are now going to use this result to study the gauge transformations and then the boundary symmetries.

\subsection{Gauge transformations}
\label{subsec:gauge}

We will now show that the generators of the gauge transformations \eqref{infinitesimal Lorentz} and \eqref{infinitesimal diffeo} are integrable and vanishing on-shell without the need to impose boundary conditions at $S$ (on either the fields or the gauge parameters). In other words, gauge transformations are degenerate directions of the extended symplectic structure even if they have support on the boundary.

Recall that the action of infinitesimal Lorentz transformations on the various fields of the extended phase space is given in \eqref{infinitesimal Lorentz} and \eqref{transformation of extra fields}. With this, we can then compute the contraction of an infinitesimal variation with the symplectic form, to find
\be\label{Gauss generator}
-\delta_\alpha\ipp\Omega\simeq-\int_\Sigma\delta e\wedge\De\alpha+\delta\omega\wedge[e,\alpha]-\int_S\alpha\delta e=-\int_\Sigma\alpha\delta(\De e)\simeq-\int_\Sigma\delta(\alpha\De e)+\L_\X(\De e)=\delta\G[\alpha],
\ee
where we have introduced the Gauss constraint
\be
\G[\alpha]\coloneqq-\int_\Sigma\alpha\De e\simeq0.
\ee
In the first on-shell equality we have discarded a term proportional to the equations of motion in the boundary integral. For the second on-shell equality, we have used the equations of motion to reintroduce a bulk term allowing us to then use formula \eqref{commuting variation and integral} in order to exchange the variation with the integral\footnote{Recall that in all such calculations of gauge generators we are allowed to use the equations of motion but not their linearized version.}. Notice that we have also assumed for simplicity that $\alpha$ is field-independent. We therefore get the announced result, namely that the generator of $\delta_\alpha$ is integrable and vanishing on-shell regardless of what happens at $S$.

The same conclusion can be reached for diffeomorphisms. At the infinitesimal level, their action on the fields of the extended phase space is given by  \eqref{infinitesimal diffeo} and \eqref{transformation of extra fields}. With this, a lengthy manipulation similar to the one above leads to
\be\label{diffeo generator}
-\L_\xi\ipp\Omega\simeq\int_\Sigma\L_\xi\omega\wedge\delta e+\L_\xi e\wedge\delta\omega-\int_S\xi\ip e\delta\omega+\xi\ip\omega\delta e\simeq\delta\D[\xi],
\ee
where we have introduced the diffeomorphism and vector constraints
\be
\D[\xi]\coloneqq\V[\xi]+\G[\xi\ip\omega]\simeq0,\q\V[\xi]\coloneqq-\int_\Sigma\xi\ip eF=-\int_\Sigma\xi\ip eF_\ell\simeq0.
\ee
Here we have chosen to write the diffeomorphism constraint as the sum of the Gauss constraint and the vector constraint in order to match what appears naturally in the canonical analysis of the theory, as we review in appendix \ref{app:usual} (where we will also write alternative expressions for $\D$). Notice that the vector constraint does not depend on the cosmological constant $\ell$ because the corresponding cubic contribution from $e$ is vanishing on the two-dimensional manifold $\Sigma$.

Now that we have studied Lorentz transformations and diffeomorphisms, we have exhausted the set of independent gauge transformations acting on our theory. However, it is still instructive at this point to discuss slightly different transformations, which are the so-called covariant diffeomorphisms (see \cite{Jacobson:2015uqa} and references therein). The interest in studying such transformations stems from the fact that the action \eqref{infinitesimal diffeo} of diffeomorphisms on the fields $e$ and $\omega$, which is defined as the usual Lie derivative, does not commute with Lorentz transformations, i.e. $[\L_\xi,\delta_{\alpha}]\neq0$. This can be cured by defining another notion of Lie derivative, the covariant Lie (or Lorentz--Lie) derivative $\L^\text{g}_\xi$, as follows:
\be
\L^\text{g}_\xi e=\De(\xi\ip e)+\xi\ip(\De e),\q\L^\text{g}_\xi\omega=\xi\ip F.
\ee
Now, recalling that the action of the standard Lie derivative \eqref{infinitesimal diffeo} can be rewritten as
\be
\L_\xi e=[e,\xi\ip\omega]+\De(\xi\ip e)+\xi\ip(\De e),\q\L_\xi\omega=\De(\xi\ip\omega)+\xi\ip F,
\ee
we can recognize that the Lie derivative and its covariant version differ by a field-dependent Lorentz transformation. In other words, we have that\footnote{Notice that in terms of Poisson brackets the action of $\delta_{\xi\ip\omega}$ on $e$ should be understood as $\lb\G[\alpha],e\rb|_{\alpha=\xi\ip\omega}$. This does however only differ from $\lb\G[\xi\ip\omega],e\rb$ by a term which vanishes on-shell.}
\be
\L^\text{g}_\xi=\L_\xi-\delta_{\xi\ip\omega},
\ee
which also implies that $\L^\text{g}_\xi u=\xi\ip\De u$ and $\L^\text{g}_\xi\ipp\X=\L_\xi\ipp\X$. In order to contract the covariant Lie derivative with the extended symplectic form, it turns out to be easier to first contract its bulk part, and then to subtract the boundary integral in \eqref{Gauss generator} for $\alpha=\xi\ip\omega$ from the boundary integral in \eqref{diffeo generator}. Doing so, we obtain that
\be
-\L^\text{g}_\xi\ipp\Omega\simeq\int_\Sigma\L^\text{g}_\xi\omega\wedge\delta e+\L^\text{g}_\xi e\wedge\delta\omega-\int_S\xi\ip e\delta\omega.
\ee
Now, we can rewrite the bulk piece as
\be\label{bulk piece of covariant diffeo}
\int_\Sigma\L^\text{g}_\xi\omega\wedge\delta e+\L^\text{g}_\xi e\wedge\delta\omega
&=\int_\Sigma\xi\ip F\wedge\delta e+\big(\De(\xi\ip e)+\xi\ip(\De e)\big)\wedge\delta\omega\nn\\
&=\int_\Sigma\xi\ip F\wedge\delta e+\xi\ip(\De e)\wedge\delta\omega-\xi\ip e\delta F+\int_S\xi\ip e\delta\omega\nn\\
&=-\int_\Sigma\xi\ip\delta eF+\xi\ip\delta\omega\De e+\xi\ip e\delta F+\int_S\xi\ip e\delta\omega\nn\\
&=-\int_\Sigma\delta(\xi\ip eF)+\xi\ip\delta\omega\De e+\int_S\xi\ip e\delta\omega\nn\\
&\simeq-\int_\Sigma\delta(\xi\ip eF)+\L_\X(\xi\ip eF)+\xi\ip\delta\omega\De e+\int_S\xi\ip e\delta\omega\nn\\
&=\delta\V[\xi]+\G[\xi\ip\delta\omega]+\int_S\xi\ip e\delta\omega,
\ee
to finally obtain that
\be
-\L^\text{g}_\xi\ipp\Omega\simeq\delta\V[\xi]+\G[\xi\ip\delta\omega]\simeq\delta\V[\xi].
\ee
We therefore get the result that the generator of the covariant diffeomorphisms is integrable on-shell, and simply given by the vector constraint $\V$.

\subsection{Boundary symmetries}
\label{subsec:symmetries}

We have seen so far how the new fields $X$ and $u$ restore gauge invariance. In particular, the generators of gauge transformations on the extended phase space are given by a bulk piece only, and as such are vanishing on-shell. As explained in \cite{Geiller:2017xad,Donnelly:2016auv,Speranza:2017gxd}, the new fields of the extended phase space happen to also support a new type of transformations on the boundary, namely the boundary symmetries. These symmetries are defined by their vanishing action on the original fields of the theory, and have an action on the new fields which is in a sense (which we will explain below) dual to that of gauge transformations. The Hamiltonian generators of these boundary symmetries form the boundary symmetry algebra.

Let us start by studying the boundary symmetries associated with Lorentz transformations. They are defined by their action on the fields of the extended phase space as
\be
\Delta_\alpha e=0,\q\Delta_\alpha\omega=0,\q\Delta_\alpha u=u\alpha,\q\Delta_\alpha X=0,
\ee
were the Lie algebra element $\alpha$ acts on $u$ from the right. This is to be contrasted with the gauge transformations $\delta_\alpha$, which act on $u$ from the left. This is the sense in which the boundary symmetries are dual to the gauge transformations, and the fact that the left and right actions commute will be key to proving that their generator is an observable. Explicitly, this generator is defined by the contraction
\be\label{variational definition Qg}
-\Delta_\alpha\ipp\Omega=\int_S\alpha(\delta\tilde e+\L_\X\tilde{e})=\int_SX_*\underline{\alpha}(\delta\tilde e+\L_\X\tilde{e})=\int_s\underline{\alpha}X^*(\delta\tilde e+\L_\X\tilde{e})=\int_s\underline{\alpha}\delta(X^*\tilde{e})=\delta\int_S\alpha\tilde{e},
\ee
where we have introduced the ``dressed'' field $\tilde{e}\coloneqq u^*e=u^{-1}eu$, and written $\alpha$ as the push-forward under $X$ of an element $\underline{\alpha}$ defined on $\mathbb{R}^3$. One can see that this expression is integrable, and therefore
\be\label{extended Gauss observable}
\Q[\alpha]\coloneqq\int_S\alpha\tilde{e}
\ee
is really the generator of the boundary symmetry $\Delta_\alpha$. Furthermore, as was already pointed out in \cite{Geiller:2017xad} (where the diffeomorphisms were however replaced by the translations), one can see that this generator is very similar to the Hamiltonian boundary charge associated with Lorentz gauge transformations on the ``usual'' phase space. It is the same expression simply with $e$ replaced by $\tilde{e}=u^*e$. Now, the symplectic structure also enables us to show that this generator is an observable. This can be done by further contracting \eqref{variational definition Qg} with the gauge transformations, which gives the following vanishing Poisson brackets:
\be
\lb\G[\alpha],\Q[\beta]\rb=-\delta_\alpha\ipp\Delta_\beta\ipp\Omega=0,\q\lb\D[\xi],\Q[\beta]\rb=-\L_\xi\ipp\Delta_\beta\ipp\Omega=0.
\ee
Finally, we can compute the algebra of the boundary symmetry generators with themselves. This is given by
\be\label{SPT LL}
\lb\Q[\alpha],\Q[\beta]\rb=-\Delta_\alpha\ipp\Delta_\beta\ipp\Omega=\Q\big[[\alpha,\beta]\big],
\ee
which is nothing but the Lorentz algebra.

We now turn to the study of the boundary symmetries associated with diffeomorphisms. While the gauge transformations which we have previously studied correspond to diffeomorphisms of the manifold $M$, the boundary symmetries correspond to diffeomorphisms $W$ on $\bb{R}^3$. As such, these diffeomorphisms do no act on the original field content of the theory, but transform the edge mode field $X$ as $X\mapsto X\circ W$. These boundary symmetries can therefore be thought of as changes of coordinates. Their action is given by
\be
\Delta_we=0,\q\Delta_w\omega=0,\q\Delta_wu=0,\q\Delta_w\ipp\X=\bbvar{W},
\ee
where $w$ denotes the vector field on $\bb{R}^3$ generating the infinitesimal version of $W$, and $\bbvar{W}\coloneqq X_*w$ is the vector field on $M$ obtained by pushing forward $w$ with $X$. With the fact that $\delta w=0$, we furthermore get the variation
\be
0=\delta w=\delta(X^*\bbvar{W})=X^*(\delta\bbvar{W}+\L_\X\bbvar{W})\q\Rightarrow\q\delta\bbvar{W}=-\L_\X\bbvar{W}=[\bbvar{W},\X].
\ee
With this, we have all the ingredients to compute the contraction of the boundary symmetry with the symplectic structure, and to analyse whether the resulting expression is integrable (i.e. whether a generator does exist). The result of the contraction is
\be\label{generic diff observable}
-\Delta_w\ipp\Omega=\delta\int_SQ[\bbvar{W}]+Q[\bbvar{W}\ip\de uu^{-1}]-\int_S\bbvar{W}\ip(\theta+\X\ip L+\de Q[\X]+\de Q[\delta uu^{-1}+\X\ip\de uu^{-1}]).
\ee
Before discussing the integrability of this expression and the existence of a generator $\Q[\bbvar{W}]$, one can already see that it will generically lead to an observable. Indeed, the Poisson bracket with the generator of Lorentz transformations is
\be
\lb\G[\alpha],\Q[\bbvar{W}]\rb=-\delta_\alpha\ipp\Delta_w\ipp\Omega=\int_S\alpha\bbvar{W}\ip(\De e)\simeq0,
\ee
and that with the generator of diffeomorphisms is
\be
\lb\D[\xi],\Q[\bbvar{W}]\rb=-\L_\xi\ipp\Delta_w\ipp\Omega=\int_S\bbvar{W}\ip\big(F_\ell\xi\ip e+\De e\xi\ip\omega\big)\simeq0,
\ee
thereby showing that $\Q[\bbvar{W}]$ (when it exists) is an observable.

In order to discuss the integrability of these generators and their algebra, we can decompose the vector field at the boundary $S$ into its tangential and normal parts as $\bbvar{W}=\bbvar{W}_\para+\bbvar{W}_\per$. Following \cite{Donnelly:2016auv} we can then call the transformations generated by vector fields with $\bbvar{W}_\per=0$ surface-preserving diffeomorphisms (SPDs hereafter), and that with $\bbvar{W}_\per\neq0$ surface translations.

It is clear from \eqref{generic diff observable} that SPDs have an integrable generator given by
\be\label{extended diff observable}
\Q[\bbvar{W}]\stackrel{\text{\tiny{SPD}}}{=}\int_SQ[\bbvar{W}]+Q[\bbvar{W}\ip\de uu^{-1}]=\int_S\bbvar{W}\ip\tilde{\omega}\tilde{e},
\ee
where for the second equality we have introduced the dressed field $\tilde{\omega}\coloneqq u^*\omega=u^{-1}\omega u+u^{-1}\de u$, thereby showing once again (just like in \cite{Geiller:2017xad} and above for Lorentz boundary symmetries) that the generator of the boundary symmetry is a dressed version of the usual Hamiltonian boundary charge. For these generators we then find the algebra
\be\label{SPT DD}
\lb\Q[\bbvar{V}],\Q[\bbvar{W}]\rb=-\Delta_v\ipp\Delta_w\ipp\Omega\stackrel{\text{\tiny{SPD}}}{=}-\Q\big[[\bbvar{V},\bbvar{W}]\big],
\ee
and with the observables associated with Lorentz transformations we have
\be\label{SPT DL}
\lb\Q[\bbvar{W}],\Q[\alpha]\rb=-\Delta_w\ipp\Delta_\alpha\ipp\Omega=\int_S\alpha\L_\bbvar{W}\tilde{e}\stackrel{\text{\tiny{SPD}}}{=}-\int_S\L_\bbvar{W}\alpha\tilde{e}=-\Q[\L_\bbvar{W}\alpha],
\ee
where we have used the fact that $\bbvar{W}$ is tangential in order to perform the integration by parts of the Lie derivative.

Together, the commutation relations \eqref{SPT LL}, \eqref{SPT DD}, and \eqref{SPT DL} show that the group of surface-preserving transformations, containing both SPDs and Lorentz transformations, is given by
\be
\text{Diff}(S)\ltimes H,
\ee
where $H=\text{SL}(2,\bb{R})$ or $H=\text{SU}(2)$ depending on the signature. This is the same result as that found in \cite{Donnelly:2016auv}. There however, there is only one generator of boundary symmetries (since in the metric formulation there are no local Lorentz transformations), given by the integral on $S$ of the Wald--Noether charge for $\bbvar{W}$. However, because in the metric formulation this charge contains derivatives of the vector field, it can naturally be decomposed into two contributions: the curvature in the normal plane to $S$ smeared with $\bbvar{W}_\para$, and the normal metric smeared with $\partial_\per\bbvar{W}_\per$. The first part has commutation relations reproducing that of $\text{Diff}(S)$, while the second one gives that of $H$ (which is $\text{SL}(2,\bb{R})$ in the Lorentzian calculation of \cite{Donnelly:2016auv}).

In order to analyse the surface translations and implement them on the extended phase space, one has to impose boundary conditions at $S$ in order to make the second term on the right-hand side of \eqref{generic diff observable} integrable. This can be done if there exists a $(2,0)$-form $B$ such that we have separately
\be\label{boundary condition B on theta with two terms}
(\theta+\de Q[\delta uu^{-1}])\big|_S=\delta B,\q(\X\ip L+\de Q[\X]+\de Q[\X\ip\de uu^{-1}])\big|_S=\L_\X B,
\ee
such that the sum is
\be\label{boundary condition B on theta}
(\theta+\X\ip L+\de Q[\X]+\de Q[\delta uu^{-1}+\X\ip\de uu^{-1}])\big|_S=\delta B+\L_\X B.
\ee
We will see below why it is necessary to have this separation between two terms. For the moment, one can notice that if this holds then the surface translations are integrable and their generator is given by
\be\label{extended diff observable + translation}
\Q[\bbvar{W}]=\int_SQ[\bbvar{W}]+Q[\bbvar{W}\ip\de uu^{-1}]-\bbvar{W}\ip B.
\ee
This condition defining $B$ can be rewritten more compactly in terms of the finite transformations of the potential. Indeed, since
\be
\theta^{X,u}=X^*\big(\theta+\L_\X\ipp\theta+\delta_{(\delta uu^{-1}+\X\ip\de uu^{-1})}\ipp\theta\big)\simeq X^*(\theta+\X\ip L+\de Q[\X]+\de Q[\delta uu^{-1}+\X\ip\de uu^{-1}]),
\ee
we can rewrite the condition of integrability of the surface translations in the form
\be
\theta^{X,u}\big|_s\simeq\delta(X^*B).
\ee
Assuming that $B$ does not depend on $X$, such that $\Delta_vB=0$, we then get that the algebra of the extended generators \eqref{extended diff observable + translation} is given by
\be
\lb\Q[\bbvar{V}],\Q[\bbvar{W}]\rb=-\Q\big[[\bbvar{V},\bbvar{W}]\big]-\int_S\bbvar{V}\ip\big(\L_\bbvar{W}B-\bbvar{W}\ip(\de B)-\de Q[\bbvar{W}]-\de Q[\bbvar{W}\ip\de uu^{-1}]\big).
\ee
By contracting the condition \eqref{boundary condition B on theta} with $\Delta_w$ one can further show that
\be
\L_\bbvar{W}B=\bbvar{W}\ip L+\de Q[\bbvar{W}]+\de Q[\bbvar{W}\ip\de uu^{-1}],
\ee
and therefore rewrite the boundary symmetry algebra in the form
\be
\lb\Q[\bbvar{V}],\Q[\bbvar{W}]\rb=-\Q\big[[\bbvar{V},\bbvar{W}]\big]-\int_S\bbvar{V}\ip\bbvar{W}\ip(L-\de B).
\ee
The remaining boundary integral on the right-hand side, although it depends on the fields, can now be shown to define a central extension of the algebra of $\text{Diff}(S)$. For this, one can compute its field space variation and show that it vanishes on-shell. The variation is given by
\be
\delta\int_S\bbvar{V}\ip\bbvar{W}\ip(L-\de B)=\int_S\bbvar{V}\ip\bbvar{W}\ip\delta(L-\de B)+\bbvar{V}\ip\bbvar{W}\ip\big(\L_\X(L-\de B)\big).
\ee
Taking the exterior derivative of the first condition in \eqref{boundary condition B on theta with two terms}, one gets that the first term is vanishing as
\be
\delta(L-\de B)\big|_S\simeq(\de\theta-\de\delta B)\big|_S=0.
\ee
Using the second condition in \eqref{boundary condition B on theta with two terms} one gets that
\be
\L_\X(L-\de B)\big|_S=\big(\de(\X\ip L)-\de(\L_\X B)\big)\big|_S=0.
\ee
Now that we have seen how the boundary condition \eqref{boundary condition B on theta} affects the Poisson bracket between the diffeomorphism boundary observables, and that it leads to a central extension, we have to also inspect the Poisson brackets with the boundary observables \eqref{extended Gauss observable}. Obviously, the commutation relations \eqref{SPT LL} are unchanged. However, because of the integration by parts of the Lie derivative, \eqref{SPT DL} becomes
\be
\lb\Q[\bbvar{W}],\Q[\alpha]\rb=\int_S\alpha\L_\bbvar{W}\tilde{e}=-\Q[\L_\bbvar{W}\alpha]+\int_S\bbvar{W}\ip\big(\de(\alpha\tilde{e})\big),
\ee
and we therefore have to understand how to interpret the extra term on the right-hand side. First of all, notice that this agrees with the other way to compute this bracket, which is
\be
\lb\Q[\alpha],\Q[\bbvar{W}]\rb=\Delta_\alpha\ipp\delta\Q[\bbvar{W}]=\Q[\L_\bbvar{W}\alpha]-\int_S\bbvar{W}\ip(\Delta_\alpha B),
\ee
since from the contraction of \eqref{boundary condition B on theta with two terms} with $\Delta_\alpha$ one can see that $\Delta_\alpha B=\de(\alpha\tilde{e})$

In order to understand the extra term in these commutation relations, we have to impose additional boundary conditions at $S$. One first possibility is to impose boundary conditions which lead to a central extension. The variation being given by
\be
\delta\int_S\bbvar{W}\ip\big(\de(\alpha\tilde{e})\big)=\int_S\bbvar{W}\ip\big(\de\big[\alpha(\delta\tilde{e}+\L_\X\tilde{e})\big]\big),
\ee
one obtains a central extension if the boundary conditions
\be
(\delta\tilde{e}+\L_\X\tilde{e})\big|_S=0
\ee
are imposed. However, one can also impose boundary conditions such that the extra term is simply vanishing (instead of being central). This can be achieved with the condition
\be
\de\Q[\delta uu^{-1}]\big|_S=\de(\delta uu^{-1}e)\big|_S=0,
\ee
which then implies that $\Delta_\alpha B=0$ and that $\lb\Q[\bbvar{W}],\Q[\alpha]\rb=-\Q[\L_\bbvar{W}\alpha]$.

\section{Conclusion}

In this article we have constructed the extended phase space containing edge mode fields for first order three-dimensional gravity with diffeomorphism and Lorentz gauge symmetry, and then studied the algebra of boundary symmetries. This is the natural continuity of previous work providing the same construction for Yang--Mills theory and metric gravity \cite{Donnelly:2016auv}, for BF and Chern--Simons theory \cite{Geiller:2017xad}, and for metric gravity with higher curvature \cite{Speranza:2017gxd}. We have followed here the systematic derivation which applies to all of these theories. This requires the introduction in the phase space of edge mode fields for each of the gauge transformations acting in the theory, namely the choice of coordinate frame $X$ in the case of diffeomorphisms, and the group elements $u$ in the case of Lorentz transformations. With these extra fields, we have constructed the extended symplectic potential \eqref{extended potential} which is invariant on-shell under finite diffeomorphisms and Lorentz transformations, and which then defines the extended symplectic structure \eqref{extended 2-form 1}. We have then studied how this extended phase space provides a natural separation between the role of gauge transformations and boundary symmetries. In section \ref{subsec:gauge}, we have shown that gauge transformations are degenerate directions of the symplectic structure even when they have support on the boundary of the local region. In section \ref{subsec:symmetries} we have defined the boundary symmetries and computed the algebra satisfied by their generators. This has revealed the same group of boundary symmetries as in the metric case, namely the semi-direct product structure $\G_S=\text{Diff}(S)\ltimes H$ in the case of surface-preserving transformations, where $H=\text{SL}(2,\bb{R})$ or $H=\text{SU}(2)$ depending on the signature. This shows that using first order variables, despite introducing local Lorentz symmetry, does not lead to an extension of the boundary symmetries. Instead, it separates the generators of $\G_S$ into two contributions: one coming from the first order diffeomorphism Hamiltonian charge, which is linear in the vector field and generates the algebra of $\text{Diff}(S)$, and one coming from the Hamiltonian charge associated with Lorentz transformations and generating the algebra of $H$ (with the bracket between the two types of generators giving the semi-direct product structure). We have also briefly discussed the boundary conditions on the extended symplectic potential which enable to promote surface translations to Hamiltonian transformations acting on the phase space.

Just as for the previous works \cite{Geiller:2017xad,Donnelly:2016auv,Speranza:2017gxd}, one important task is now to understand the representation theory of the boundary symmetries (this is already partly understood for the Ka\v c--Moody algebras appearing in BF and Chern--Simons theory \cite{Bimonte:1992xe}), and how it can be used to regularize and define entanglement entropy in the gravitational context. In three dimensions, we expect that the availability of different descriptions of the boundary symmetries (namely as $\G_S$, with possible central extensions, or as the Ka\v c--Moody algebras of BF and Chern--Simons theory) will shed light on this problem. The availability of this dual description means essentially that one can trade the effects of diffeomorphisms for that of internal symmetry transformations whose representation theory is in principle easier to construct.

Finally, one important and interesting line of work would be to understand the relationship between the various symmetry groups and algebras which are known to appear in three-dimensional gravity, and which describe spacetime symmetries or the physics of point particles. These depend of course on the signature of spacetime, on the sign of the cosmological constant, on the location of the boundary (at finite distance or infinity) and on possible boundary conditions or choice of asymptotic behavior for the fields. To be more concrete, let us consider the case of a vanishing cosmological constant. Then, the group of symmetries preserving the structure of asymptotically-flat spacetime is the three-dimensional BMS group BMS$_3$ \cite{Bondi:1962px,Sachs:1962zza,Sachs:1962wk,Ashtekar:1978zz,Ashtekar:2014zsa,Grumiller:2017sjh}. Representations of BMS$_3$ were studied in \cite{Oblak:2016eij,Barnich:2014kra,Barnich:2015uva}, where it was also shown that one can rewrite $\text{BMS}_3=\text{Diff}(S)\ltimes\text{Vect}(S)$. This makes it explicit that BMS$_3$ is actually ``smaller'' than the group $\G_S$ of boundary symmetries at finite distance which we have derived in this paper. However, this remarks begs for a clearer definition, as BMS$_3$ and $\G_S$ are not defined at the same location in spacetime and in terms of the same geometrical structures, and it is not clear how the former is embedded in the latter. In addition, it is known that BMS$_3$ admits certain natural and physically relevant central extensions \cite{Oblak:2016eij}, and it would therefore be interesting to compare these to the central extensions discussed in the main text and to that appearing in the Ka\v c--Moody boundary algebra of BF and Chern--Simons theory. Finally, the fact that we have here studied finite boundaries is reminiscent of structures appearing when coupling point particles to three-dimensional gravity. Indeed, this coupling typically requires the introduction of small boundaries around the particles in order to regularize the curvature and torsion contraints \cite{Buffenoir:2003zu}. In the quantum theory, the mass and spin of point particles are representation labels of an algebra which is not that of the classical spacetime, but which presents a quantum deformation. This is given by the Drinfel'd double of $\text{SU}(2)$ in the flat Euclidean case for example \cite{Bais:2002ye,Bais:1998yn,Noui:2006ku}. Now, it turns out that representations of BMS$_3$ do also lead to a notion of particles, which are however not strictly classical particles but instead BMS$_3$ particles dressed by soft modes. We leave the precise study of these mathematical structures for future work, hoping that it will help clarify further the role and the physical interpretation of the boundary symmetries at finite distance which we have described in this work.

\section*{Acknowledgements}

I would like to thank William Donnelly, Laurent Freidel, Wolfgang Wieland and Gabriel Wong for discussions. This research is supported by Perimeter Institute for Theoretical Physics. Research at Perimeter Institute is supported by the Government of Canada through the Department of Innovation, Science and Economic Development, and by the Province of Ontario through the Ministry of Research \& Innovation.

\appendix

\section{Relation to usual boundary observables}
\label{app:usual}

As explained in \cite{Geiller:2017xad} and seen in the main text, the boundary observables which generate the boundary symmetries on the extended phase space are a dressed version of the ``usual'' Hamiltonian boundary observables defined on the non-extended phase space. In this appendix we recall the construction of these usual Hamiltonian boundary observables following \cite{Regge:1974zd,Balachandran:1994up,Balachandran:1995qa,Husain:1997fm} and using the criterion of functional differentiability of the constraints. It should be kept in mind that the calculations of this appendix do \textit{not} use the edge mode fields $X$ and $u$ and the extended phase space.

We start with the $2+1$ decomposition of the action, which will identify the constraints of the theory. This is given by
\be
S
&=\int_Me\wedge\left(F+\f{1}{6\ell^2}[e\wedge e]\right)\nn\\
&=\int_\bb{R}\de t\int_\Sigma\de^2x\,\teps^{ab}\left(\partial_0\omega_ae_b+\omega_0D_ae_b+e_0\left(\f{1}{2}F_{ab}+\f{1}{2\ell^2}[e_a,e_b]\right)-\partial_a(e_b\omega_0)\right)\nn\\
&=\int_\bb{R}\de t\int_\Sigma\partial_0\omega\wedge e+\omega_0\De e+\f{1}{\sqrt{q}}NEF_\ell+M\ip eF-\de(e\omega_0),
\ee
where we have introduced the density vector
\be
E^i\coloneqq\f{1}{2}\teps^{ab}[e_a,e_b]^i=[e_1,e_2]^i,\q E^iE^j\eta_{ij}=q\coloneqq\det(q_{ab}),\q E^ie^j_a\eta_{ij}=0,
\ee
and used the change of variables
\be
e_0^i=M^ae_a^i+\f{1}{\sqrt{q}}NE^i,\q N\coloneqq\f{1}{\sqrt{q}}e_0^iE^j\eta_{ij},\q M^a\coloneqq\f{1}{q}\teps^{ab}E^i[e_0,e_b]^j\eta_{ij}.
\ee
This rewriting of the triad Lagrange multiplier is what enables to trade the curvature constraint $F_\ell$ (enforced by $e_0$ and generating the translations \eqref{infinitesimal translations}) for the scalar and vector constraints. The smeared constraints of the theory are then given by the following Gauss, scalar, and vector constraints:
\be
\G[\alpha]=-\int_\Sigma\alpha\De e,\q\C[\N]=-\int_\Sigma\f{1}{\sqrt{q}}\N EF_\ell,\q\V[\xi]=-\int_\Sigma\xi\ip eF_\ell=-\int_\Sigma\xi\ip eF,
\ee
for a Lie algebra-valued function $\alpha$, a scalar $\N$, and a vector field $\xi$. The vector constraint does not depend on the cosmological constant because the corresponding cubic contribution from $e$ is vanishing on the two-dimensional manifold $\Sigma$.

Let us now compute the variation of the constraints and at the condition of functional differentiability. Assuming that $\delta\alpha=0$, the variation of the Gauss constraint is given by
\be
\delta\G[\alpha]=\int_\Sigma\De\alpha\wedge\delta e+[e,\alpha]\wedge\delta\omega-\int_S\alpha\delta e.
\ee
This can be made functionally differentiable without restricting the dynamical fields at $S$ by using smearing fields $\bar{\alpha}$ vanishing\footnote{We will denote with an overline all the smearing fields which are compactly supported, i.e. vanishing at $S$.} at $S$. Then, for an arbitrary $\alpha$ which does not have to vanish at $S$ the quantity
\be\label{usual Gauss observable}
\Q[\alpha]=\int_\Sigma\De\alpha\wedge e=\G[\alpha]+\int_S\alpha e\simeq\int_S\alpha e
\ee
is an observable since
\be
\lb\G[\bar{\alpha}],\Q[\beta]\rb=\G\big[[\bar{\alpha},\beta]\big]+\int_S[\bar{\alpha},\beta]e=\G\big[[\bar{\alpha},\beta]\big]\simeq0,
\ee
and it satisfies the Poisson brackets
\be\label{usual LL algebra}
\lb\Q[\alpha],\Q[\beta]\rb=\Q\big[[\alpha,\beta]\big].
\ee
Notice that these brackets are well-defined without additional conditions since $\Q[\alpha]$ is functionally differentiable if $\delta\alpha=0$. One can see from this calculation that the procedure leading to the Hamiltonian observable \eqref{usual Gauss observable} is to integrate the Gauss constraint by parts, discard the boundary term, and use an arbitrary smearing field $\alpha$ which is not constrained to vanish at $S$. This then commutes with the constraint since this latter is well-defined only for smearing fields $\bar{\alpha}$ that are vanishing at $S$.

Similarly, the scalar constraint $\C[\N]$ can be made functionally differentiable without restricting the dynamical fields at $S$ by choosing smearing functions $\bar{\N}$ vanishing at $S$. One can then attempt at constructing an observable by integrating by parts, discarding the boundary term, and using an arbitrary $\N$. However, it is easy to see that the resulting quantity is again not functionally differentiable unless this a priori arbitrary smearing field is also constrained to vanish at $S$. This therefore shows that there are no Hamiltonian observables arising from the scalar constraint via this construction. Of course, this is the case because we are here insisting on not restricting the dynamical fields at $S$, and only the smearing fields. Once again, recall that the reason for studying this condition is that this is what reproduces the observables which we have constructed on the extended phase space.

Instead of studying next the vector constraint, we are going to study the constraint generating spatial diffeomorphisms. This is a combination of the vector and Gauss constraints given by the following equivalent expressions:
\be
\D[\xi]
&=\V[\xi]+\G[\xi\ip\omega]\nn\\
&=-\int_\Sigma\de\omega\xi\ip e+\de e\xi\ip\omega\nn\\
&=\int_\Sigma\big(\xi\ip(\de\omega)\big)\wedge e+\big(\xi\ip(\de e)\big)\wedge\omega\nn\\
&=\int_\Sigma\L_\xi\omega\wedge e-\int_S\xi\ip\omega e\nn\\
&=\int_\Sigma\L_\xi e\wedge\omega-\int_S\xi\ip e\omega.
\ee
Assuming that $\delta\xi=0$ the variation of this constraint is given by
\be
\delta\D[\xi]=\int_\Sigma\L_\xi\omega\wedge\delta e+\L_\xi e\wedge\delta\omega-\int_S\xi\ip e\delta\omega+\xi\ip\omega\delta e.
\ee
Again, this can be made functionally differentiable without restricting the dynamical fields at $S$ by using smearing vector fields $\bar{\xi}$ that are vanishing at $S$, and then one indeed has that
\be
\lb\D[\bar{\xi}],e\rb=\L_{\bar{\xi}}e,\q\lb\D[\bar{\xi}],\omega\rb=\L_{\bar{\xi}}\omega.
\ee
Let us now consider the quantity
\be\label{usual diff observable}
\Q[\xi]=\int_\Sigma\L_\xi\omega\wedge e=\D[\xi]+\int_S\xi\ip\omega e\simeq\int_S\xi\ip\omega e
\ee
for an arbitrary $\xi$ which does not have to vanish at $S$. This corresponds indeed to the diffeomorphism constraint once we have integrated by parts and discarded the boundary term. Since
\be
\delta\Q[\xi]=\int_\Sigma\L_\xi\omega\wedge\delta e+\L_\xi e\wedge\delta\omega+\int_S\xi\ip(\delta\omega\wedge e),
\ee
this can be made functionally differentiable without restricting the dynamical fields if $\xi$ is restricted to be tangential to $S$. The quantity $\Q[\xi]$ is then an observable since
\be
\lb\D[\bar{\xi}],\Q[\zeta]\rb
=\int_\Sigma\L_\zeta\omega\wedge\L_{\bar{\xi}}e-\L_{\bar{\xi}}\omega\wedge\L_\zeta e
=-\D\big[[\bar{\xi},\zeta]\big]\simeq0,
\ee
and the algebra of these observables is given by
\be\label{usual DD algebra}
\lb\Q[\xi],\Q[\zeta]\rb
=-\Q\big[[\xi,\zeta]\big].
\ee
In addition, we have that the Poisson bracket with the observables \eqref{usual Gauss observable} is given by
\be\label{usual DL algebra}
\lb\Q[\xi],\Q[\alpha]\rb=-\Q[\L_\xi\alpha].
\ee

As expected, the algebra defined by \eqref{usual LL algebra}, \eqref{usual DD algebra} and \eqref{usual DL algebra}, is isomorphic to the algebra $\text{Diff}(S)\ltimes H$ of surface preserving symmetries defined in the main text. In addition, the observable \eqref{extended Gauss observable} on the extended phase space is a dressed version of the observable \eqref{usual Gauss observable}, and the observable \eqref{extended diff observable} is a dressed version of \eqref{usual diff observable}. Finally, while on the usual phase space the construction of these observable requires to constrain the smearing fields at $S$, this requirement is relaxed on the extended phase space.

Now, it is also the case that the usual boundary observables derived above correspond to the Hamiltonian boundary charges on the non-extended phase space, where the symplectic structure has no boundary contribution and is simply given by
\be\label{bulk symplectic structure}
\Omega_\Sigma=-\int_\Sigma\delta\omega\wedge\delta e.
\ee
We recall here these calculations in order to show in particular that the generator of the gauge-covariant diffeomorphisms is in general non-integrable even in the case of tangential diffeomorphisms at $S$. For Lorentz transformations we have
\be
-\delta_\alpha\ipp\Omega_\Sigma=-\int_\Sigma\delta e\wedge\De\alpha+\delta\omega\wedge[e,\alpha]=\delta\G[\alpha]+\int_S\alpha\delta e,
\ee
so the generator is integrable if $\delta\alpha=0$ but not vanishing on-shell. Instead, it is equal to the Hamiltonian boundary charge. For diffeomorphisms we have
\be
-\L_\xi\ipp\Omega_\Sigma=\int_\Sigma\L_\xi\omega\wedge\delta e+\L_\xi e\wedge\delta\omega=\delta\D[\xi]+\int_S\delta\omega\xi\ip e+\delta e\xi\ip\omega,
\ee
and the surface term is equal to
\be\label{charge for diffeos}
\int_S\delta Q[\xi]-\xi\ip\theta=\int_S\delta(e\xi\ip\omega)-\xi\ip(\delta\omega\wedge e),
\ee
which is therefore integrable if $\xi$ is tangential at $S$. Finally, for gauge-covariant diffeomorphisms we can repeat the calculation \eqref{bulk piece of covariant diffeo} (without the weak equality since on the non-extended phase space there are no edge mode fields $X$) to find
\be
-\L^\text{g}_\xi\ipp\Omega_\Sigma=\int_\Sigma-\delta(\xi\ip eF)-\xi\ip\delta\omega\De e+\int_S\xi\ip e\delta\omega=\delta\V[\xi]+\G[\xi\ip\delta\omega]+\int_S\xi\ip e\delta\omega,
\ee
which shows that the gauge-covariant diffeomorphisms are not integrable (even in the case of tangential diffeomorphisms).

\section{Gibbons--Hawking--York corner term}
\label{appendix:GHY}

In this appendix we compute the Hamiltonian charges for Lorentz transformations, diffeomorphisms and covariant diffeomorphisms, in the case where the symplectic structure is extended on the boundary $S$ by using the corner contribution coming from the Gibbons--Hawking--York (GHY) boundary term. It should be kept in mind that the calculations of this appendix do \textit{not} use the edge mode fields $X$ and $u$ and the extended phase space constructed in the main text.

In first order variables, the GHY boundary term can be constructed in terms of the internal unit normal $n^i=n^\mu e^i_\mu$. The action then takes the form \cite{0264-9381-4-5-011,Bodendorfer:2013hla,Corichi:2015nea,Wieland:2017zkf}
\be
S[e,\omega,n]=\int_Me\wedge\left(F+\f{1}{6\ell^2}[e\wedge e]\right)+\int_{\partial M}[e,n]\wedge\De n,
\ee
where $\De n=\de n+[\omega,n]$ and $n^in^j\eta_{ij}=\pm1$ depending on the signature. Treating the normal $n$ as a dynamical variable, the variation of the boundary term is given by
\be\label{variation of GHY}
\delta([e,n]\wedge\De n)=\delta[e,n]\wedge\De n+[e,n]\wedge\delta[\omega,n]+\de[e,n]\delta n-\de([e,n]\delta n),
\ee
and one can see that the integration by part produces a corner contribution. As explained in \cite{Geiller:2017xad}, if the boundary of the manifold is smooth then the boundary terms and their corner contributions do not contribute to the symplectic structure. However, if the total boundary is not smooth one can consider a mixed variational principle where boundary terms and corner contributions can be added separately on (say) the time-like boundary. This is for example how boundary contributions to the symplectic structure were added in \cite{Wieland:2017zkf,Freidel:2015gpa,Freidel:2016bxd}. Here one can use the same reasoning to add the corner contribution from the GHY boundary term to the potential, and thereby obtain a boundary contribution to the symplectic structure. Thinking carefully about sings and orientations shows that one should actually add the corner piece of \eqref{variation of GHY} to the potential with a minus sign, so that the total symplectic structure becomes $\Omega=\Omega_\Sigma+\Omega_S$, with $\Omega_\Sigma$ the bulk symplectic structure \eqref{bulk symplectic structure} used repeatedly above, and
\be
\Omega_S=\int_S\delta[e,n]\delta n=\int_S[\delta e,n]\delta n+[\delta n,\delta n]e.
\ee

With this boundary symplectic structure, one can then compute the contraction with infinitesimal Lorentz transformations and diffeomorphisms to obtain
\be
-\delta_\alpha\ipp\Omega_S=-\int_S\alpha\delta e,\q-\L_\xi\ipp\Omega_S=\int_S\delta([n,\de n]\xi\ip e)-\xi\ip(\de e)[n,\delta n]\stackrel{\text{\tiny{SPD}}}{=}\int_S\delta([n,\de n]\xi\ip e),
\ee
where we have assumed that the diffeomorphism vector field $\xi$ is tangential at $S$. With this it is then immediate to see that
\be
-\delta_\alpha\ipp\Omega=\delta\G[\alpha],
\ee
which means that Lorentz transformations have a vanishing charge. For (tangential) diffeomorphism we get that
\be
-\L_\xi\ipp\Omega\stackrel{\text{\tiny{SPD}}}{=}\delta\D[\xi]+\delta\int_S\xi\ip e(\omega+[n,\de n]),
\ee
and using $\xi\ip e\omega=[\xi\ip e,n][\omega,n]$ the charge can be rewritten as
\be
\int_S[\xi\ip e,n]\De n.
\ee
Finally, the covariant diffeomorphisms now become integrable as well and we have
\be
-\L^\text{g}_\xi\ipp\Omega\stackrel{\text{\tiny{SPD}}}{=}\delta\V[\xi]+\G[\xi\ip\delta\omega]+\delta\int_S[\xi\ip e,n]\De n,
\ee
so their charge is the same as that of non-covariant diffeomorphisms. A little rewriting now shows that this charge is nothing but the Komar charge which we would have obtained for diffeomorphisms in the metric formulation, i.e.
\be\label{Komar charge}
\int_S[\xi\ip e,n]\De n\simeq\int_S(k_an_b-n_ak_b)\nabla^a\xi^b.
\ee 
The final result of this appendix is therefore that the algebra of boundary charges is unchanged if we introduce the normal $n$ and the corner contribution from the GHY term. The effect of this extension of the phase space including $n$ is to set the charges of Lorentz transformations to zero, but this information about Lorentz transformations is in a sense ``transferred'' to the diffeomorphisms, whose charges then do not take the first order form \eqref{charge for diffeos}, but instead become the Komar expression \eqref{Komar charge}. Since the result of the present work was to show that the algebra of Lorentz and first order diffeomorphism charges (whether it is realized on the extended or on the usual phase space) is the same as that of second order (or metric) diffeomorphism Komar charges, we see that the inclusion of $n$ in the phase space via the GHY corner term does also leave this algebra unchanged.

\section{Chern--Simons theory}

For the sake of completeness, and because of the known and important relationship between Chern--Simons theory and three-dimensional gravity, we discuss in this appendix the extended phase space of Chern--Simons theory with edge mode fields associated with internal gauge transformations and diffeomorphisms.

The case of internal gauge transformations was treated briefly in \cite{Geiller:2017xad}, and we recall here the details. The Lagrangian and the potential are given by
\be
L=A\wedge\left(F-\f{1}{6}[A\wedge A]\right),\q\theta=\delta A\wedge A.
\ee
The extended pre-symplectic potential which is gauge-invariant on-shell under finite gauge transformations $g\in G$ is
\be
\theta_\text{e}\coloneqq\theta+2\de(A\delta uu^{-1})+\delta(A\wedge\de uu^{-1})+\de(\delta uu^{-1})\wedge\de uu^{-1},
\ee
where the $G$-valued edge mode field $u$ transforms as $g^*u=g^{-1}u$. To compute the extended symplectic structure, we now use the fact that
\be
\delta\big(\de(\delta uu^{-1})\wedge\de uu^{-1}\big)=\de\big(\de(\delta uu^{-1})\delta uu^{-1}\big),
\ee
and obtain $\Omega=\Omega_\Sigma+\Omega_S$, with the usual bulk symplectic structure
\be\label{bulk CS symplectic structure}
\Omega_\Sigma=-\int_\Sigma\delta A\wedge\delta A,
\ee
and the boundary symplectic structure
\be
\Omega_S=\int_S\big(2\delta A+\De(\delta uu^{-1})\big)\delta uu^{-1}.
\ee
We can now contract infinitesimal gauge transformations (which act on $u$ as $\delta_\alpha u=-\alpha u$) with the extended symplectic structure to find
\be\label{CS gauge contraction}
-\delta_\alpha\ipp\Omega=2\int_\Sigma\De\alpha\wedge\delta A-2\int_S\alpha\delta A=\delta\F[\alpha],
\ee
were
\be
\F[\alpha]=-2\int_\Sigma\alpha F\simeq0
\ee
is the flatness constraint. Boundary symmetries $\Delta_\alpha$ are now defined by their action $\Delta_\alpha A=0$ on the gauge field and $\Delta_\alpha u=u\alpha$ on the edge mode field. With this, we get the contraction
\be
-\Delta_\alpha\ipp\Omega=2\int_Su\alpha u^{-1}\big(\delta A+\De(\delta uu^{-1})\big)=2\delta\int_S\alpha\tilde{A},
\ee
where the dressed gauge field is $\tilde{A}\coloneqq u^*A=u^{-1}Au+u^{-1}\de u$. The generator is therefore integrable and given by
\be
\Q[\alpha]=2\int_S\alpha\tilde{A}.
\ee
As expected this is gauge-invariant, i.e.
\be
\lb\F[\alpha],\Q[\beta]\rb=-\delta_\alpha\ipp\Delta_\beta\ipp\Omega=0,
\ee
and leads to the centrally-extended algebra
\be
\lb\Q[\alpha],\Q[\beta]\rb=2\int_S\De(u\alpha u^{-1})u\beta u^{-1}=\Q\big[[\alpha,\beta]\big]+2\int_S\de\alpha\beta.
\ee
This is the Ka\v c--Moody algebra of the Lie algebra of $G$.

Let us now discuss diffeomorphisms. These are in general realized in Chern--Simons theory as field dependent gauge transformations with parameter $\alpha=\xi\ip A$. Because of this field dependency, on the usual (i.e. non-extended) phase space carrying the symplectic structure \eqref{bulk CS symplectic structure} the Hamiltonian charge is not integrable unless additional boundary conditions are specified. Here one can however see that on the extended phase space the contraction \eqref{CS gauge contraction} becomes
\be
-\delta_{\xi\ip A}\ipp\Omega\simeq2\int_\Sigma\De({\xi\ip A})\wedge\delta A-\delta(\xi\ip A)F-2\int_S{\xi\ip A}\delta A=\delta\F[\xi\ip A],
\ee
where we have used the equations of motion to introduce an extra curvature term. This shows indeed that $\F[\xi\ip A]$ is the on-shell generator of diffeomorphisms.

Another way to study diffeomorphisms is to follow the approach of the main text and to introduce edge mode fields $X$. For Chern--Simons theory, the Noether current and charge associated with diffeomorphisms are given by
\be
J
&=\L_\xi A\wedge A-\xi\ip L\nn\\
&=\De(\xi\ip A)\wedge A+\xi\ip F\wedge A-\xi\ip L\nn\\
&=\de(\xi\ip AA)-\xi\ip A\De A+\xi\ip F\wedge A-\xi\ip L\nn\\
&=\de(\xi\ip AA)-2\xi\ip AF\nn\\
&\simeq\de(\xi\ip AA).
\ee
The extended symplectic structure can be found following the same reasoning as in the main text, and it is therefore given by \eqref{extended 2-form 1} with $Q[\X]=\X\ip AA$ and $u=\mathds{1}$, which is
\be
\Omega=\int_\Sigma\delta\theta+\int_S\X\ip\theta+\f{1}{2}\X\ip\X\ip L+\delta A\X\ip A-\f{1}{2}A[\X,\X]\ip A-A\X\ip\delta A+\X\ip\big(\de(A\X\ip A)\big).
\ee
With this we get that
\be
-\L_\xi\ipp\Omega\simeq2\int_\Sigma\L_\xi A\wedge\delta A-2\int_S\xi\ip A\delta A\simeq\delta\D[\xi],
\ee
with
\be
\D[\xi]\coloneqq-2\int_S\de A\xi\ip A.
\ee
For boundary symmetries, we have
\be
-\Delta_w\ipp\Omega=\delta\int_SQ[\bbvar{W}]-\int_S\bbvar{W}\ip(\theta+\X\ip L+\de Q[\X]),
\ee
which is of course identical to the result of the main text with $u=\mathds{1}$ and $Q[\bbvar{W}]$ the diffeomorphism Noether charge computed above.

\bibliography{Biblio}
\bibliographystyle{JHEP}

\end{document}